# SOFTWARE SECURITY ANALYSIS DYNAMIC AND STATIC BUFFER CODE ANALYSIS WITHIN JAVA AND C PROGRAMMING ENVIRONMENT A COMPARATIVE STUDY

MANAS GAUR

AMBEDKAR INSTITUTE OF TECHNOLOGY, INDIA

1. EXECUTIVE SUMMARY

*1.1 purpose*

The main stretch in the paper is buffer overflow anomaly occurring in major source codes, designed in various programming language. It describes the various as to how to improve your code and increase its strength to withstand security theft occurring at vulnerable areas in the code.

The main language used is JAVA, regarded as one of the most object oriented language still create lot of error like stack overflow, illegal/inappropriate method overriding. I used tools confined to JAVA to test as how weak points in the code can be rectified before compiled. The byte code theft is difficult to be conquered, so it's a better to get rid of it in the plain java code itself. The tools used in the research are PMD(Programming mistake detector), it helps to detect line of code that make pop out error in near future like defect in hashcode(memory maps) overriding due to which the java code will not function correctly. Another tool is FindBUGS which provide the tester of the code to analyze the weak points in the code like infinite loop, unsynchronized wait, deadlock situation, null referring and dereferencing. Another tool which provides the base to above tools is JaCoCo code coverage analysis used to detect unreachable part and unused conditions of the code which improves the space complexity and helps in easy clarification of errors.

1.2 *Audience*

Through this paper, we design an algorithm to prevent the loss of data. The main audience is the white box tester who might leave out essential line of code like, index variables, infinite loop, and inappropriate hashcode in the major source program. This algorithm serves to reduce the damage in case of buffer overflow.

1.3 *Scope*

We have performed testing of near about 4200 LOC in java of static buffer overflow and behavior of above mentioned tools towards buffer overflow. We also lay the comaparison between static techniques and dynamic techniques towards buffer overflow. We also tested the applications server JBOSS and buffer overflow theft in legacy code like string processing , overriding etc. We also establish an efficient ways to program a code to be free from buffer overflow.

*Keywords:- stack overflow, buffer overflow, code performance, hascode analysis, statement coverage*

2. INTRODUCTION

(The size and complexity of software systems is growing , increasing the number of
bugs.According to Carnegie Mellon University, CERT, the number of of vulnerabilities
have increase with nearly 500% in 2years as shown
.

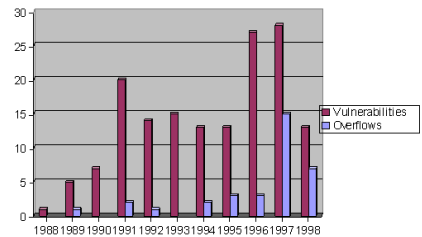

Figure 1

**Buffer Overflow** has been the Major threat for security vulnerability in the last ten years.
Moreover, buffer overflow vulnerabilities dominate in the area of remote network
penetration vulnerabilities,where an anonymous Internet user seeks to gain partialor total
control of a host. Because these kinds of attacks enable anyone to take total control of a
host, they represent one of the
 most serious classes security threats. . There are many security breaches in the universe but have there root as susceptible code,
open data or use of static structures in code like array etc or incomplete hashing of the source.

**The U.S. is the country most affected,** as they provide about 80% of the world's software. Software counterfeiting is claimed to
be a large problem by some, resulting in a revenue loss of US $11-12 billion, China and Vietnam are the biggest offenders.
Furthermore, The Fourth Annual BSA and IDC Global Software Piracy Study revealed that 35% of the software installed in 2006
on personal computers (PCs) worldwide was obtained illegally, amounting to nearly $40 billion in global losses due to software.
The rate of copyright infringement of software in the Asia-Pacific region has been estimated at 53% for 200.[US LAW REPORT
by ANIRUDH RAO]

**LEGACY CODE:** Many programs developed in the past generally on any programming language might have used its libraries
for its development which might have deprecated in the coming years, leading the code vulnerable to threats.

```
public
class Thread implements Runnable {
...
@Deprecated
   public final void stop() {
  synchronized (this) {
...
```

The stops method in the above code has been deprecated. Generally String processing code which uses Strcpy(), Strcat(), String
Buffer() class and String Builder() class /interface does not provide the complete implementation of the method of there
respective classes /interface faces threats like:-
Pointers deallocation
Change of address of pointers
Deprecated code of strings eg:
NULL pointer exception

EXPERIMENTAL SET UP
In this paper we investigates the effectiveness of 2 publicly available tools for static prevention of **buffer overflow** in java codes
,1 tool for dynamic testing java code and 1 tool that helps in detecting the coverage of the code and helps to extract out
untouched and unreachable part of the code.

   3.1 *Findbugs*
         **FindBugs** is a smart tool used in detecting static intrusion points in the code. Findbugs is generally applied in static
testing of java source programs and provides the data defining the priority of error, confidence factor, type of error and its effect
on other part of the module.
      3.1.1 *Bug pattern in Findbugs*
         Malicious code vulnerability:- code that can be altered by other code. Dodgy :- code that can lead to error.Bad
Practice:- code that violates the recommended coding practice. Correctness:- code that might give different results than the
developer intended. Internationlization:- code that can inhibit the use of international characters. Performance- code can be
transformed to provide better performance. Security:- security problems in the code. Multithreaded correctness- multithreaded
environment threats. Experimental- no closing statement of streams, database objects or others require closing statements.
      3.1.2 *Features of Findbugs*
     a. **Bug Explorer Window**:- It tells briefly the type of the bug
generated on the run. It also tells how many bugs generated and  how
many are filtered.

     b. **Properties Window**:- It tells about the description of the
error, its priority and also the marker, parent and child relationship in the
code.                                  Figure 2
     c. **Bug info window**(also embeds bug navigator window):-

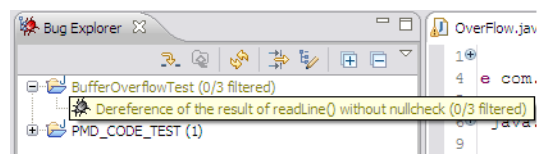

Figure 2

provides complete description of the error, with its category, rank, confidence and type.
    **Category:-** which bug pattern is followed in the code eg:- dodgy, malicious. **Rank:-** tells the effect of error i;e RANK=15 then error is of concern. **Confidence:-** level of error. **Type:-** whether error is null pointer, deadlock or stack overflow.

### 3.1.3 Disadvantages
The disadvantages of these tools is they don't understand what your software is trying to do, and their sense of context is extremely limited which can lead to a lot of false positives being reported which developers have to waste their time reviewing.

## 3.2 PMD(Programming mistake detector)

**Programming Mistake Detector** is a static code analyzer for Java. Developers use PMD to comply with coding standards and deliver quality code. Team leaders and Quality Assurance folks use it to change the nature of code reviews. PMD has the potential to transform a mechanical and syntax check oriented code review into a to dynamic peer-to-peer discussion.

**Figure 3**

### 3.2.1 What is PMD?
PMD works by scanning Java code and checks for violations in three major areas

#### 3.2.1.1 Compliance with coding standards such as:
Naming conventions - class, method, parameter and variable names
Class and method length
Existence and formatting of comments and JavaDocs

#### 3.2.1.2 Coding antipatterns such as:
Empty try/catch/finally/switch blocks
Unused local variables, parameters and private methods
Empty if/while statements
Overcomplicated expressions - unnecessary if statements, for loops that could be while loops
Classes with high Cyclomatic Complexity measurements

#### 3.2.1.3 Cut and Paste Detector (CPD)
- a tool that scans files and looks for suspect code replication. CPD can be parameterized by the minimum size of the code block.[9]

**Important**:- PMD comes with 149 rules and 19 ruleset but it also provides the tester to develop his own set of rules to test the code and to bring homgenity in code .

### 3.2.2 Priority assignment in PMD
PMD assigns violation priority from 1 to 5. They are as follows
VERY HIGH PRIORITY(indicated with red)
HIGH PRIORITY(indicated with orange)
MEDIUM PRIORITY(indicated with yellow)
IGNORANT PRIORITY(indicated with green)
NEGLIGIBLE(indicated with blue)

Finally, in the **PMD Violations** table it is possible to add review annotations to the source where the violation occurred. This can be an effective way to mark target files for further review. Just right-click on any violation in the list and select **Mark review**. PMD will insert a review annotation to the Java source right above the violation line itself. The review annotation should look like this:

   //**@PMD:REVIEWED: MethodNamingConvention: by MAK GAUR on 3/28/12 12.04PM**

### 3.2.3 PMD works
PMD relies on the concept of **Abstract Syntax Tree (AST),** a finite, labeled tree where nodes represent the operators and the edges represent the operands of the operators. PMD creates the AST of the source file checked and executes each rule against that tree. The violations are collected and presented in a report. PMD executes the following steps when invoked from Eclipse (Based on PMD's documentation). The Eclipse PMD plugin passes a file name, a directory or a project to the core PMD engine (Also part of the Eclipse plugin, however housed in a separate package). This engine then uses the RuleSets as defined in the PMD preferences page to check the file(s) for violations. In the case of a directory or project (multiple source files) the plugin executes the following steps for each file in the set.
PMD uses JavaCC to obtain a Java language parser.PMD passes an InputStream of the source file to the parser.The parser returns a reference of an Abstract Syntax Tree back to the PMD plugin.PMD hands the AST off to the symbol table layer which builds scopes, finds declarations, and find usages.If any rules need data flow analysis, PMD hands the AST over to the DFA layer for building control flow graphs and data flow nodes.Each Rule in the RuleSet gets to traverse the AST and check for violations.The

Report is generated based on a list of RuleViolations. These are displayed in the PMD Violations view or get logged in an XML, TXT, CSV or HTML report.[9]

Figure 4

### 3.2.4 Abstract Syntax tree in PMD

In computer science, an **abstract syntax tree** (AST), or just **syntax tree**, is a tree representation of the abstract syntactic structure of source code written in a programming language. Each node of the tree denotes a construct occurring in the source code. The syntax is 'abstract' in the sense that it does not represent every detail that appears in the real syntax. For instance, grouping parentheses are implicit in the tree structure, and a syntactic construct such as an if-condition-then expression may be denoted by a single node with two branches. Abstract syntax trees are also used in program analysis and program transformation systems.

## 3.3 JaCOCO by Eclemma

JaCoCo uses a set of different counters to calculate coverage metrics. All these counters are derived from information contained in Java class files which basically are Java byte code instructions and debug information optionally embedded in class files. This approach allows efficient on-the-fly instrumentation and analysis of applications even when no source code is available. In most cases the collected information can be mapped back to source code and visualized down to line level granularity.

### 3.3.1 Instructions (C0 Coverage)

The smallest unit JaCoCo counts are single Java byte code instructions. **Instruction coverage** provides information about the amount of code that has been executed or missed. This metric is completely independent from source formatting and always available, even in absence of debug information in the class files

### 3.3.2 Branches (C1 Coverage)

JaCoCo also calculates *branch coverage* for all if and switch statements. This metric counts the total number of such branches in a method and determines the number of executed or missed branches. Branch coverage is always available, even in absence of debug information in the class files. Note that exception handling is not considered as branches in the context of this counter definition.

No coverage: No branches in the line has been executed (red diamond)
Partial coverage: Only a part of the branches in the line have been executed (yellow diamond)
Full coverage: All branches in the line have been executed (green diamond)

### 3.3.3 Cyclomatic Complexity

JaCoCo also calculates **cyclomatic complexity** for each non-abstract method and summarizes complexity for classes, packages and groups. According to its definition by McCabe in 1996 cyclomatic complexity is the minimum number of paths that can, in (linear) combination, generate all possible paths through a method.[7] Thus the complexity value can serve as an indication for the number of unit test cases to fully cover a certain piece of software. Complexity figures can always be calculated, even in absence of debug information in the class files.

**The formal definition of the cyclomatic complexity** $v(G)$ is based on the representation of a method's control flow graph as a directed graph:

$v(G) = E - N + 2$

**Where E is the number of edges and N the number of nodes**. JaCoCo calculates cyclomatic complexity of a method with the following equivalent equation based on the number of branches (B) and the number of decision points (D):

$v(G) = B - D + 1$

Based on the coverage status of each branch JaCoCo also calculates covered and missed complexity for each method.[8] Missed complexity again is an indication for the number of test cases missing to fully cover a module. Note that as JaCoCo does not consider exception handling as branches try/catch blocks will also not increase complexity.

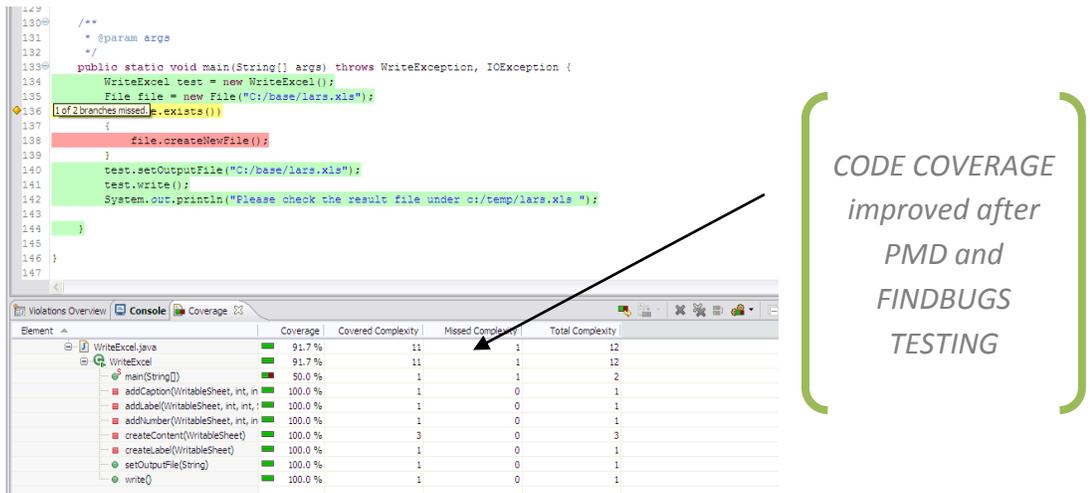

Figure 5

### 3.3.4 JaCoCO works

#### 3.3.4.1 Lines

For all class files that have been compiled with debug information, coverage information for individual lines can be calculated. A source line is considered executed when at least one instruction that is assigned to this line has been executed. Due to the fact that a single line typically compiles to multiple byte code instructions the source code highlighting shows three different status for each line containing source code:

No coverage: No instruction in the line has been executed (red background)
Partial coverage: Only a part of the instruction in the line have been executed (yellow background)
Full coverage: All instructions in the line have been executed (green background)

#### 3.3.4.2 Methods

Each non-abstract method contains at least one instruction. A method is considered as executed when at least one instruction has been executed. As JaCoCo works on byte code level also constructors and static initializers are counted as methods.[11] Some of these methods may not have a direct correspondence in Java source code, like implicit and thus generated default constructors or initializers for constants.

#### 3.3.4.3 Classes

A class is considered as executed when at least one of its methods has been executed. Note that JaCoCo considers constructors as well as static initializers as methods. As Java interface types may contain static initializers such interfaces are also considered as executable classes.

### 3.4 Asertn

**Asertn** tool that is developed as part of this research. The tool enables he programmer to add the assertion atatements and analyzes the behhaviour of the agent at run time to compare it against the intent of the agent programmer.

3.4.1 The **Lexical Analyzer** will break the code into tokens.
3.4.2 The **Parser** will convert Java tokens into a parse tree where as assertion statements will be represented as a list that consists of the assertion tokens and an identifier of the corresponding scope labels.[wikipeadia]
3.4.3 **Test code generator** (instrumentor): converts the assert statements to Java statements and inserts them in the relevant locations in the Java source code.
3.4.4 **The Java run-time environment** consisting of (Compiler, Linker, and Executer) is responsible for executing the asserted source code that is expressed in Java Language.
3.4.5 **After execution the tool** will read the collected information and presents it as a report to the user describing for each error the name of the program unit that has the error, the statement.

## 4. ANALYSIS

### 4.1 Bug patterns

1. Infinite recursive Loop: this is one of serious coding error not detected by compiler but can make your influential to attacker. When a function is called the caller and its address are put to stack and if the called function again makes a call to caller/itself its again puts it address on the stack , if there is no exit the stack overflows and the code is dead.
   Example:-
   Public static void main(String args[])
   {makeover();}
   Void makeover()
   {Makeoverdone();}
   Void Makeoverdone()
   {If(1)
   Makeover(); // not EXIT and the condition always point to true and the call goes on loop and stack
      a. // overflows.}

2. Hashcode and Equals
   Java.lang. super class files and default equals method which can be called as it is and does not errupt an error. But since it is a default one you cannot force it to behave according to you, for that you have to override it. This is the step where major errors originate because with every "EQUALS" there is associated hashcode, which is used for hashing( memory management in the operating system) the bytecode of a compilede java code.
   Example:Public static void main(String args[]) {
      a. String str="hello"; String str1="HELLO"
      b. If( str.equals(str1)) Print(" they are equal");}
   The equals called is the defaut one and sometime the generated output is malignant. Therefore we override the equals method with the hashcode. If the hashcode is not defined the compiler will generate is default hashcode and the data and variables will be stored in some anonymous location in the memory.The correct implementation is
   //same code above Public int equals(String st){
   If (str == str1) Return 1;
   Else Return 0;} Public int hashcode()
   { Assert false: " string is errorneous";
   Return 434; }

3. Null pointer dereferencing

   **public static void** main(String args[]) **throws** FileNotFoundException{

   a. CODE 1:       File file = **new** File("C:\\Documents and Settings\\MAK\\workspace\\BufferOverflowTest\\src\\com\\BOT1\\text1.txt");
      i. FileReader isr= **new** FileReader(file);
      ii. BufferedReader br = **new** BufferedReader(isr);
      iii. **int** newvals = 0;
      iv. **double**[] nvals = **new double**[10];

         1. **double**[] vals = **new double**[10];
         2. **try**{ **while**(br.readLine()!= **null**)
         3. {String str = br.readLine();

      v. *vals[newvals]=Double.valueOf(str.trim()).doubleValue(); //bug detected*

         1. newvals++;}nvals=vals;}
      vi. **catch**(IOException ioe){System.*out*.checkError(); }

4. Return values ignored :- this means when the called function returns a value to the caller to pop up from stack the caller should check the value as it might create security breach.
5. Inconsistent Synchronisation:- In todays programming  areana every one wants multiprogramming and multithreading, in java it is obtained by wait, sleep, join and the synchronized block. The code involving the above keywords should use conditional block like
6. If-else{}, switch-case{} and at a time only one object has the right to execute as it might lead to deadlock or UNCONDITIONAL WAIT SITUATION  thereby eating away memory spaces leading to low system utilization.

### 4.2 Static Analysis

Findbugs Size of the analyzed codes

2,115 LOC and 56 Java Classes

Findbugs Results :- the total time that Findbugs used to analyze the code and warning was 9.3 seconds. we define the results of findbugs in 3 category.

Relevant True Positive- a bug developer should fix.

Irrelevant Positive-not of bug, but can be in future.

False positive- not a bug of concern

#### 4.2.1 Experimental results of code 1

Information about the bug [experimental results on Eclipse IDE indigo] :

**Bug**: Dereference of the result of readLine() without nullcheck in.
com.BOT1.OverFlowFile.main(String[])The result of invoking readLine() is dereferenced without checking to see if the result is null. If there are no more lines of text to read, readLine() will return null and dereferencing that will generate a null pointer exception. **Confidence**: Normal, **Rank**: Of Concern (15)

**Pattern**: NP_DEREFERENCE_OF_READLINE_VALUE

**Type**: NP, **Category**: STYLE (Dodgy code)

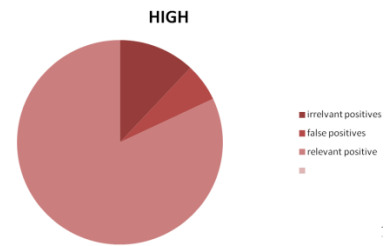

Figure 6

#### 4.2.2 Relevant True Positives

The cases under this section relates to code that should be changed to eliminate error.

Associated Code:- *public class CheckersGameState implements State{*
*public static int BOARD_SIZE=8;}*

*Error Description*

**The following description was attached to the error:**

**aialg.checkers.CheckersGameState.BOARD_SIZE** isn't final but should be
CheckersGameState.java Group10 Game Framework/src/aialg/checkers line 16

**FindBugs Problem (High Priority)** The error from FindBugs indicates that BOARD_SIZE should be final since it is a static variable that is being assigned a constant value. This is a valid defect since not declaring BOARD_SIZE final allows the variable to be changed which can cause problems in other code where BOARD_SIZE used and assumed to be unchanging.

#### 4.2.3 Irrelevant true Positives

**public class Building** {private Lift lift; private int floor; public Building(){ *lift = new Lift();*
setFloor(15); } public int getFloor(){ return floor; } public void setFloor(int floor){ this.floor = floor; } }
**public class Lift**{ private Building building;private int Speed;
public Lift(){*building = new Building();* setSpeed();} public void setSpeed(){if
(building.getFloor()>20){ this.Speed = 10; }else {this.Speed = 5;}}
public int getSpeed(){return Speed}}}

ERROR DESCRIPTION:- FindBugs problem (Normal Priority)

The above code has *circular dependency* , it should be avoided as it introduces unnecessary dependencies. However removing it requires refactoring of the code, with findbug priority tells that we can leave the code.

#### 4.2.3 False Positives

False Positive were cases where changing the code to eliminate errors reported by the tool would be incorrect. The tool reported three false positives which were different instances of the error. These false positives may be valid error in production code, but were invalid in the context of the test code.

Experimental results:- public class RobustnessTest {
 public Game createInitializationBoardWithPlayers() throws Exception {
  Game testCheckers = null;
  CheckersGameFactory testFactory = new CheckersGameFactory();
  testCheckers = testFactory.createGame();
  *assertTrue(testCheckers != null);*
  Player testWhite = testCheckers.createPlayer("testWhite", Color.white);
  Player testBlack = testCheckers.createPlayer("testBlack", Color.black);
  *assertTrue(testWhite != null);*
  *assertTrue(testBlack != null);* testCheckers.addPlayer(testWhite);

testCheckers.addPlayer(testBlack); testCheckers.initialize();   return testCheckers;  } }
public class CheckersGameFactory extends GameFactory {
public Game createGame() { return new CheckersGame(); } }
**Error Description**

The following description was attached to the first of the three errors (with similar descriptions for the other two errors:
**Redundant nullcheck of testCheckers, which is known to be non-null
RobustnessTest.java Group10 Game Framework/src/tests line 47
FindBugs Problem (Low Priority).**

4.3 *PMD(Programming Mistake Detector)*
  4.3.1 Experimental Results
      CODE 2:  **public class** Yang **extends** Thread {
           **public** Yang(String str) {
           **super**(str);}
           **public void** run() {
           **for** (**int** i = 0; i < 10; i++) {
         System.*out*.println(i + " " + getName());
           **try** {*sleep*((**long**) (Math.*random*() * 1000));} **catch** (InterruptedException e) {}}
         System.*out*.println("DONE! " + getName());}
           **public void** WRITE_SOMETHING(String INPUT_PARAMETER) {
           System.*out*.println(INPUT_PARAMETER);}

           **public static void** main(String args[] ){**new** Yang("Good").start();**new** Yang("Bad").start();}

           **public** String thisIsCutAndPaste(String pFirst, **int** pSecond) {
           System.*out*.println("New world"); **return** "New world";}}

The results shows the efficiency of PMD test on a yang.java code, its respose to bufferoverflow exceptions. It also tells the error per method which set apart PMD from Findbugs.
PMD has a unique feature of setting your own exception to make a coherent code which is unambiguous and ambiguous in syntax formation.

Figure 7

4.4 *JaCoCo by Eclemma Code Coverage Analysis*
 4.4.1 Postive Results

Figure 8
  4.4.2 Negative Results(Missed Branches)

Figure 9

The snippet shown above displays the missed branch in the code line;-

   *If(!file.exists())* this is a conditional test, creating unreachable branch making code vulnerable to threat. Such results can be tested otherwise via a complexity graph and tracing a minimum spanning tree, which helps in sorting out unused nodes, which can lead unintended results and software malfunctioning.

     4.4.3  Cyclomatic Complexity

        The result shown above is of an Excel Project created by us.

       4.4.3.1 Explanation Of Results

Coverage:- It is calculated via expression

Figure 10

(No. Of line(Covered & missed)/ Total lines Covered)*100

Covered Complexity:- this entity tells the lines reached/reachable at compile/runtime. If its value equals to Total Lines Covered, means each branch is reachable and code is free from threat.

Missed Complexity:- defines the branches(LOC) not reachable during testing. If its value equals total LOC the code is vague and nearer to security theft.

Total Complexity:- derived by drawing a graph nodes(index variables, conditional statement) and edge(flow of data) . it is calculated by:= E(Edges)-N(Nodes) +2 and traversing a graph by dijkastra or warshall algo to derive transitive closure , the 0 values in matrix shows missed branches.

$$\begin{pmatrix} 1 & 1 & 1 \\ 1 & 1 & 1 \\ 1 & 1 & 1 \end{pmatrix} 3 \times 3$$

    4.5  Exe File /.class file Analysis

       When we compile code it in nessarily converted in .class file which is generally in low level, more esoteric for the operating system. A developer can never rectify errors like heap overflow, stack overflow(though can provide exception handling in code but its internal to code) neither integer overflow. A Overflow situation generally occurs when we use array implementation or size limit in arraylist like datastructures, we have generated some experimental results on stackoverflow, heap overflow or integeroverflow.String processing code are ,more vulnerable so we need to impose strong constraint on the siz e of the stack(forC/C++ user use Strncat instead of strcat, strncmp for strcmp etc.).

**EXAMPLE 1**

| STACK ADDRESS | VALUE |
|---|---|
| 0000 | 0049 |
| O004 | 0088 |

**EXAMPLE 2**

| STACK ADDRESS | VALUE |
|---|---|
| 0000 | 0066 |
| 0004 | 1234 |

**EXAMPLE 3**

| STACK ADDRESS | VALUE |
|---|---|
| 0000 | 9999 |
| 0004 | 9999 |

The values in **Example 1 represent** the location in the program where execution will resume when the current function completes its tasks and a value passed to a called function. In other words, a function was called during program execution. In order for the program to know where to resume once the called function returns control back to the calling function, the address of the next line of code to be executed is stored in the stack. In this case, the value of that address is 0049. The value of 10 in offset 0004 represents a value passed to the called function. In this example, there's no problem. The programmer assumed that the value passed to the called function would not exceed 4 bytes. The called function executes, and control is returned to the appropriate line of code in the calling function. [4]

In **Example 2**, an attacker is taking advantage of a buffer overflow vulnerability in the application. The attacker found that the programmer didn't add code to verify the size or data type of the data passed to the called function. By entering the value 12340088 (which exceeds the expected 4-byte limit), the attacker has succeeded in overwriting the return address stored in offset 0000 with the address of a malicious executable. When the called function completes its tasks, control will be handed over to the malicious program at address 0088 instead of the next line in the calling function at address 0049.
Not all buffer overflow attacks are designed to cause the execution of malicious code.
In **Example 3**, the attacker simply entered a series of 9's. In this case, the program will probably crash when it attempts to return control to the calling function.[4]
The data provided to the called function might come from a variety of sources. The key point to take from this example is that the input was not properly validate. (Note: For you purists out there, I know that certain values in the stack might not be stored most significant digit first. This is just easier for demonstration purposes.)

**Heap Overflow**
When a program retrieves a large amount of data for processing, a portion of memory known as the heap is allocated to handle the loaded data. In low-level languages like C and C++, the programmer is responsible for ensuring the proper amount of memory is set aside. If the loaded data is larger than the allocated heap memory, the system could crash.[10]

**Integer Overflow**
When adding two integers, the result occasionally exceeds the memory allocated for the result.
When added together, the following two eight bit integers (10 + 5) fit nicely into an eight bit result space:
0000 1010 (10)
+0000 0101 ( 5)
0000 1111 (15)
But what about the following:
1100 0000 (208)
1101 0000 (192)
0001 1001 0000 (400)
The sum of 400 won't fit in an 8 bit memory space.
The integer overflow is not necessarily a good vehicle for outside attacks. But if your application doesn't return an exception error, your data integrity might be a little off. In this case, you might end up with a value of 144 (1001 0000) instead of 400 in your database or in your next processing step.

*4.6 Asertn Tool Analysis*
        4.6.1 Code Test
                Asertn tool uses the concept of assertion, which is a conditional testing and can     be enabled or disabled at run time.

```
  Public static void main(String args[]){
BufferedReader br= new BufferedReader(new InputStreamReader(System.in));
String str=br.readLine(); // not size definition provided
Int a = Integer.parseInt(str);// susceptible ares of bugs
Assert(a>10): System.out.println("its false");
/rest code here}
```
Assertion in java can be used for dynamic testing as we can enable or disable the assertion block at the compilation or testing time and can test behavior of working of the code.

Example
C:\sun\sdk\bin> javac –ea filename.java
<press enter>
This will enable the assertion block. By default assertion is closed.

## 5. RESULTS

5.1 *Algorithm For Buffer Overflow Bound Checking*

| |
|---|
| **FINDING ARRAY.** |
|     1.1 UnderLine each array declaration |
|     1.2 For Each array underline all subsequent reference. |
| **INDEX VARIABLES:-** legal ranges of an array of size n is 0<i<N |
|     2.1 For each underlined access that uses a variable as an index write legal range next to it |
|     2.2 For each index marked in 2.1 underline all occurrences of that variable. |
|     2.3 Circle any assignements , input or operation that may modify these index variable. |
|     2.4 Mark with a V(or any letter) that is indexed by an circled index variable. |
| **LOOPS THAT MODIFY INDEX VARIABLE.** |
| 3.1 Find loops that modify variables used to index arrays. |
| 3.2 For any index that occurs as part of the loop conditional, underline the loop limit.<br>    For example: - for(i=0; i<max; i++) if I is the index variable underline i<max. |
| 3.3 Write the legal range of the array index next to the loop limit as you did in 2.1. Mark a V if the loop limit could exceed the legal range of the array index. |
| 3.4 Watch out for the loop that goes until i<=max as the largest valid index is max-1. |
| 3.5 If the upper or lower loop limit is a variable, it must be declared , it must be checked just as indices are checked in step 2. |

*Comparative Analysis*
Java Tools For Evaluation

Table 1

| TOOLS | ANALYSIS STRATEGY | EFFICIENCY RATE |
|---|---|---|
| FINDBUGS | Static analysis, flow sensitive analysis, java bytecodes | 42.4% |
| PMD | Unused variables, symmetricity in code, error in exception handling, garbage collection | 51.8% |
| JaCoCo(Code Coverage) | Date flow,Complexity analysis | 17.98% |
| Asertn | Dynamic Analysis, assertion error, error handling | 13.81% |

C/ C++ Tools Efficiency rate

Table 2

| TOOLS | EFFICIENCY RATE |
|---|---|
| BOON(Buffer Overrun detectiON) | 9.09% |
| SPLINT(Secure Programming Lint) | 15.528% |
| ARCHER(Array Checker) | 74.4% |

Note:- ALL tools in comparison are static analysis tools.
Efficiency rate is Buffer overflow detection rate.
Calculated as
(TOTAL NO. OF WARNINGS(positive**)** / Total Lines Of Code) **X 100**

5.2.3 Comparison between FindBugs And PMD

Table 3

| TYPES OF ERROR | FINDBUGS | PMD |
|---|---|---|

| Concurrency warning | 72 | 4 |
|---|---|---|
| Null Dereferncing | 27 | 0 |
| Null Assignment | 6 | 68 |
| Index Out Of Bounds | 19 | 11 |

Note:- Data Collected out of 320 LOC of my project on ONLINE EXAM in JAVA

### 5.2.4 Different Tools Find Different Bugs

Table 4

| BUG CATEGORY | EXAMPLE | FINDBUGS | PMD |
|---|---|---|---|
| General | Null Dereference | YES | YES |
| Concurrency | Possible DeadLock | YES | YES |
| Array | Length may be zero | YES | NO |
| Conditional Loop | Unreachable code | YES | YES |
| String Processing | Check equality(==,!=) | YES | YES |
| Object Overriding | HASCODE check | YES | YES |
| IO Stream | Streams closed or not | YES | NO |
| DESIGN | Static inner classes | YES | NO |
| Unnecessary stmt. | Ignored return statement | NO | YES |

### 5.2.5 Proof By Code

```
Import java.io.*;
Public class Testing{Private byte[] b; Pivate int size;
Testing(){size=25;
         b=new byte[size];}
     Public void test(){int z; // Variable unused detected by PMD
     try{FileInputStream fis-new FileInputStream("XYZ");
      x.read(b,0,size); // Method value ignored detected by FindBugs
      c.close();}
    catch(Exception e)
    // IO stream unclosed on exception caught detected by     Findbugs
    {System.out.println("help, I m caught");}
    for(int y=1;  y<=size; y++)
   {If(Integer.toString(50)==Byte.toString(b[i]))
   //Using == for comparing string detected by findbugs
    System.out.println(b[i] + "");}}//end of test method}//end of class
```

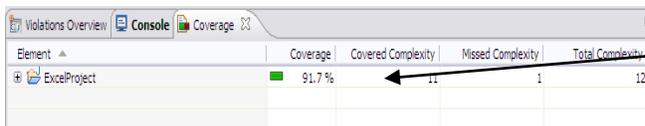

*Coverage with 91.7 % After PMD*

Figure 11

## 6. LITERATURE SURVEY

### 6.1 Static/ Dynamic Intrusion Prevention Tools

#### 6.1.1 Libsafe and Libverify

These tools are the combination of static and dynamic interusion prevention. The key  idea behind Libsafe is to estimate a safe boundary for buffers on the stack at run time and then check this boundary before any vulnerable function is allowed to write. LibVerify is an enchancement , it involves return address verification and since it uses library functions , doesn't require recompilation.[libsafe][1][2]

Vulnerable functions in C.\/C++ are

Strcpy(char *d, char * src) : overflow with pointer d: repairmen, use strncpy(char *d, char * src, int size);
Strcat(char *d, char *src) : overflow with pointer d: repairement, use strncat(same syntax above)
gets(char *s) : overflow with pointer s: define the size for the pointer or use conditional statements.

#### 6.1.2 Purify

Rational® Purify® is a dynamic software analysis tool designed to help developers write more reliable code. It includes two capabilities: 1) Memory debugging (pinpoints hard to find memory errors such as uninitialized memory access, buffer overflow and improperfreeing of memory), 2) Memory leak detection (identifies memory blocks that no longer have a valid pointer). Purify is supported on Windows®, Linux®, Solaris®, and AIX®.[IBM.com]

### 6.1.3 Verisoft(UNIX testing tool)

**Benefits.** VeriSoft can *quickly* reveal behaviors that are virtually impossible to detect using conventional testing techniques, and hence *reduces the cost* of *testing and debugging*, while *increasing reliability*.

**Scope.** VeriSoft can test software applications developed in *any language* (C, C++, Tcl, etc.). VeriSoft is optimized for analyzing *multi-process* applications. It can analyze systems composed of processes described by hundreds of thousands of lines of code. Source code for all the components is not required.

**Technology.** The key technology used in VeriSoft is a new form of *systematic state-space exploration* (also called *"model checking"* in the research literature). With its first prototype developed in 1996 and its design first presented at POPL'97, VeriSoft is the *first* software model checker using a run-time scheduler.

### 6.2 Related Works

Three other studies of defenses against buffer overflow attacks have been made

1. **Web Application security –Buffer overflows Are you really at risk by TOM OLZAK in 2006**, The research says about the various different types of overflows that happens in system when a code is put to test and also provide a comparison between JAVA and other programming language but does not provide a valid stable proof for the comparison. It's a kind of universal research not pertaining to any language which doesnot include JAVA's FINDBUGS approach and CODE coverage analysis without which testing is half done.
2. A comparison of publicly available tools for dynamic buffer overflow prevention by JOHN WILANDER and MARIAM KAMKAR. The research is quiet detailed covering every aspect in C/C++ using UNIX but there is flaw not of great priority is that the research is restricted ton one OS and One programming language. Its not a comparative research and proof is not is very explanatory. A Comparative study helps in selection effective language for development.reserach about deprecated files is not involved and for dynamic analysis you need have an alogorthm , not written in this paper.
3. In late 2000 Crispin Cowan published there paper"buffer overflow : Attak and defences for the Vulnerability of the decade. They implicitly discuss several of our attack forms but leave out the integer overflow and data structures overflow. Comparison of defenses is broader considering also OS, choice of programming language and code auditing, but there is only theoretical analysis , not comparative and discussed limited static and dynamic tools.

## 7. CONCLUSION

FindBugs and PMD analysis of the code really lowers the threat of the software. With the tremendous rise in object oriented programming the threats increase, at syntax level, bytecode and unused part of the code which consumes only memory. We put forward an algorithm that rectify the erroneous lines and perform bounded buffer checking. The coverage analysis of an untested code is merely 21.7 % but after static analysis we can achiece about 91 % coverage. Large unreachable lines in program bring in security breach like buffer overflow and diversion of the program from its intended use. We conclude , findbugs and PMD analysis at static level and assertions at dynamic level retards the fragile lines in program and reduces its space complexity.

## 8. FUTURE DIRECTIONS

### 8.1 *Future Scope in Findbugs*

The fact that Findbugs support only java, limits its uses to java based application. However findbugs support detection of various categories of bugs like;-
Performance bugs in embedded applications
Concurrency bugs in Complex multithreading.
Priority Based analysis of Bugs( +200 bugs detection in findbugs).

### 8.2 *Future Scope in PMD*

In our research we focused on static buffer overflow ,coherent code generation and violation detection using PMD but it has far more scope in future in areas like:-
Data Flow Analysis
Better Symbol Analysis
Code Cleanup- detecting and correcting sloppy codes.

### 8.3 *Future Scope of JaCOCO*

We used this tool for analyzing cyclomatic complexity of code ,indetifying statement coverage and branch coverage, however this tool serves a pancea for various white box testing and finds its utility in detecting missing requirement in software engineering and Uml diagrams. It can also serves as tool for Feasibility analysis of software projects.